\documentclass[3p,times,twocolumn]{elsarticle}
\usepackage{amssymb}
\usepackage[figuresright]{rotating}

\begin{document}

\begin{frontmatter}



%vl\dochead{Regular article}

\title{High-rate glass MRPC for fixed target experiments at Nuclotron}


\author[1]{N.A.~Kuzmin}
\author[1]{E.A.~Ladygin}
\author[1]{V.P.~Ladygin}%\corref{cor1}}
%%\cortext[cor1]{vvvvv}
\author[1]{Yu.P.~Petukhov}
\author[1]{S.Ya.~Sychkov\fnref{label3}}
\fntext[label3]{deseased} 
\author[2]{A.A.~Semak}
\author[2]{M.N.~Ukhanov}
\author[3]{E.A.~Usenko}
\address[1]{Joint Institute for Nuclear Research, Dubna, Russian Federation}
\address[2]{ Institute for High Energy Physics, Protvino, Russian Federation}
\address[3]{Institute for Nuclear Research, Russian Academy of Sciences, Moscow, Russian
Federation}

\begin{abstract}
A Multi-gap Resistive Plate Chamber (MRPC) equipped  with heaters to improve the  counting rate capability was designed for the BM@N  experiment in Dubna.
The measurements were performed using a muon beam at IHEP U-70 accelerator in Protvino. 
The MRPC at 40$^\circ$C tolerates counting rate up to 6 kHz/cm$^2$ with time resolution $\sim$65 ps and efficiency  $\sim$95\%
which complies with the conditions of the experiment.
\end{abstract}


\end{frontmatter}

%\linenumbers

~
%\vskip 1cm

\section{Introduction}
\label{intro}

Modern experiments in high energy physics and relativistic heavy ion collisions require good
particle identification based on the time-of-flight (TOF) techniques.
Multi-gap Resistive Plate Chamber, first developed in
1996 \cite{mrpc_1996}, is a gas detector with good time resolution and high detection efficiency.
Further developments of MRPCs %\cite{mrpc_2000,mrpc_alice1,mrpc_alice2}
\cite{mrpc_2000}-\cite{mrpc_alice2}  
gave a possibility to build   large area TOF systems. 
Nowadays MRPCs are used for particle identification at
 ALICE \cite{mrpc_alice3,mrpc_alice4}, CMS \cite{mrpc_cms}, 
STAR \cite{mrpc_star1}-\cite{mrpc_star3},  
%\cite{mrpc_star1,mrpc_star2,mrpc_star3},
PHENIX \cite{mrpc_phenix}, 
HARP \cite{mrpc_harp1,mrpc_harp2},  
FOPI \cite{mrpc_FOPI1, mrpc_FOPI2}, HADES \cite{mrpc_Hades1}-\cite{mrpc_Hades3} 
%\cite{mrpc_Hades1,mrpc_Hades2,mrpc_Hades3}
and other experiments.

New generation of fixed target experiments in relativistic heavy ion collisions 
investigating highly compressed baryonic matter  at moderate temperatures, CBM at FAIR \cite{cbm} and BM@N at
NICA \cite{bmn1}-\cite{bmn3}, %\cite{bmn1,bmn2,bmn3}, 
require
good particle identification  in the conditions of
high charged particle flux.
For instance, the fluxes in the inner zones of the CBM and BM@N TOF systems
are expected to be up to $\sim$25 kHz/cm$^2$ \cite{cbm_tof} and  $\sim$5 kHz/cm$^2$ \cite{bmn2}, respectively.    
However, counting rate of MRPC made with conventional float glass with a bulk resistivity in
the range  10$^{12}$ -- 10$^{13}$ $\Omega$cm is limited to several hundreds Hz/cm$^2$.
Therefore, the extension of the counting rate capabilities of MRPC has become an important issue.

To increase the MRPCs performance at high rates a low resistivity  
glass (less than 10$^{10}$ -- 10$^{11}$ $\Omega$cm) 
%\cite{mrpc_phglass, ammosov1, china1, mrpc_lhc} 
\cite{mrpc_phglass}-\cite{mrpc_lhc}
or ceramic \cite{mrpc_ceramics} can be used as the electrode materials.  
For instance,  time resolutions below 90 ps and efficiency higher than 90\%
were obtained for fluxes up to 25 kHz/cm$^2$ for the 10-gap MRPC
\cite{china1}. Another method to increase MRPC rate capability is to reduce the electrode
thickness and to increase temperature of the glass
\cite{heat1, heat2}. Such method can provide good time resolution 
at the rate up to $\sim$20 kHz/cm$^2$ \cite{gapienko1}. 
It was demonstrated recently that a moderate warming (up to 40$^\circ$C) can
enhance the detector rate capability \cite{nim_cern}.

The flux in BM@N experiment is not continuous \cite{bmn1,bmn2}.
It increases sharply during the slow extraction of the beam from Nuclotron ($\sim$3 s),
while it is negligible during the rest of the acceleration cycle ($\sim$10 s). 
This regime differs significantly from the one with the continuous flux on the
MRPC \cite{mrpc_lhc}.  The time resolution and the efficiency
should be studied as a function of time during the beam spill. 

The paper presents the measurements of 12-gap MRPC performance at the
counting rate ranging from 0.45 kHz/cm$^2$ up to 10 kHz/cm$^2$ and  temperature  
range from 25$^\circ$C to 45$^\circ$C. The data were taken using a muon beam at the U70 accelerator in Protvino \cite{U70_status}.
The paper is organized as following. Chapter \ref{design} describes
the design of the MRPCs.
The associated front-end electronics (FEE) is  discussed in chapter \ref{fee}.
The beam test at U70 accelerator and detail of the data  analysis are described in chapter \ref{U70}.   
The results are discussed in chapter  \ref{results}.
The conclusions are drawn in the last chapter. 

\section{Design of MRPC}
\label{design}

\begin{figure}[hbtp]
\centering
\includegraphics*[width=80mm,height=50mm]{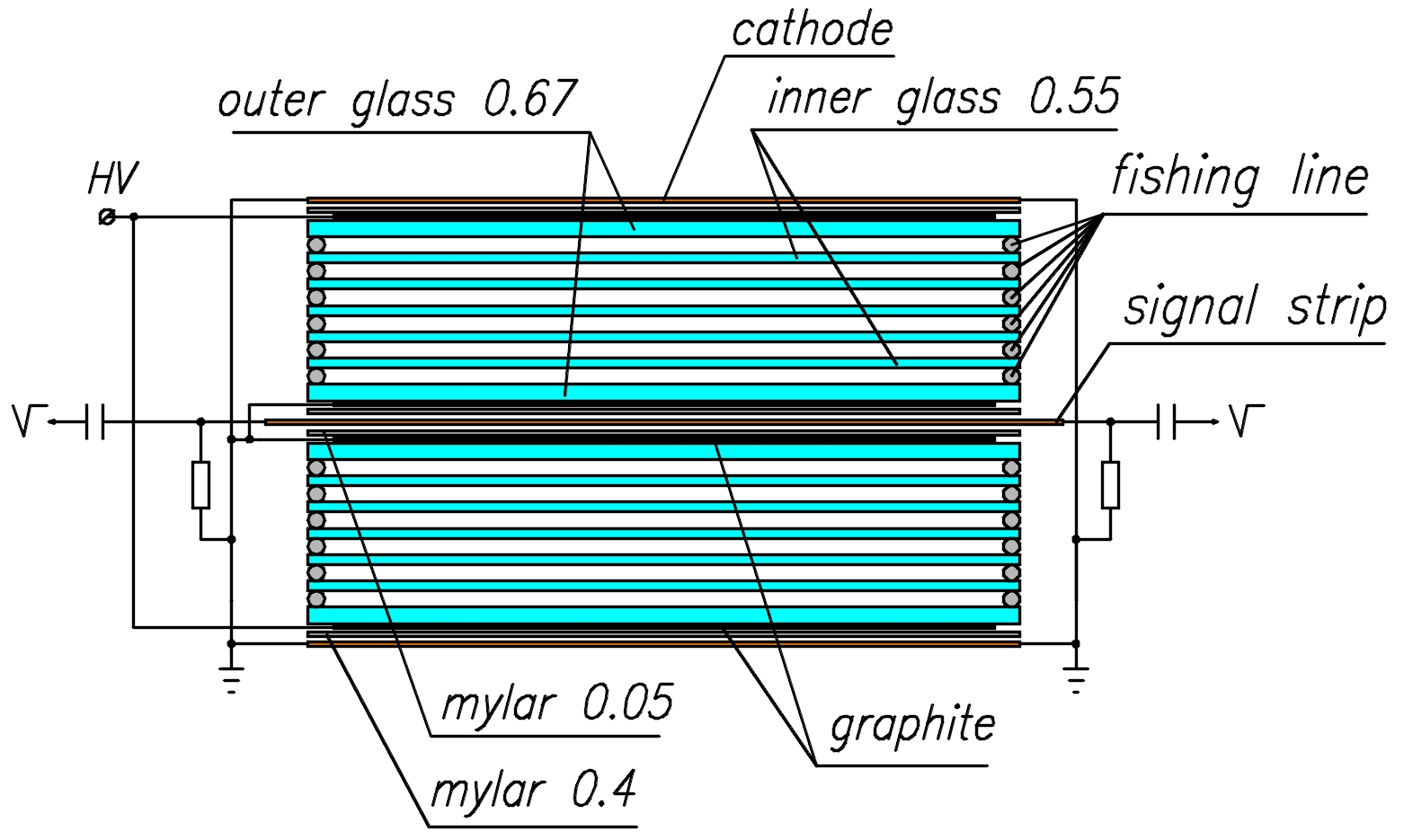}
\caption
{The schematic view of the MRPC.
}
\label{fig:fig1}
\end{figure}

Schematic cross section of MRPC is shown in Fig.\ref{fig:fig1}. The  MRPC consists of 
two identical 6-gap stacks (12 gas gaps in total)
 with anode strip readout plate in between.
The size of MRPC is 473$\times$ 279$\times$17 mm$^3$ with the sensitive area  of
351$\times$160 mm$^2$.  The MRPC has 32 10$\times$160 mm$^2$ readout strips  
with 1 mm gaps between them.

Bulk resistivity of the glass plates is 
2$\cdot$10$^{12}$ $\Omega$cm,  
thickness is 0.55~mm.
The gap between the glass plates is defined by a spacer made of a fishing-line 0.22~mm in diameter. 
The  thickness of the outer glass plates in the stack is 0.67~mm.
Mylar film with a thickness of 0.4~mm is placed between the cathode and outer glass plate.  
Graphite conductive coating with the surface resistivity of $\sim$1 
M$\Omega$/square is painted on outer surfaces of the stacks to distribute 
high voltage to create an electric field in the sensitive area.
The anode readout strips plate is a one-sided printed PCB with the thickness 
of 100~$\mu$m.  The thickness of the copper coating is 35~$\mu$m. 
The signals are read on both sides of the anode strips.
The PCB is insulated from the outer electrode by a  50~$\mu$m thick Mylar layer.
The entire MRPC is enclosed in a gas-tight aluminium box. The bottom of the box is made of a double-sided PCB (motherboard) with a thickness of 2.5~mm. The top of the box is covered by an aluminium plate 1.5~mm thick. 
 
\begin{figure}[hbtp]
\centering
\includegraphics*[width=70mm,height=60mm]{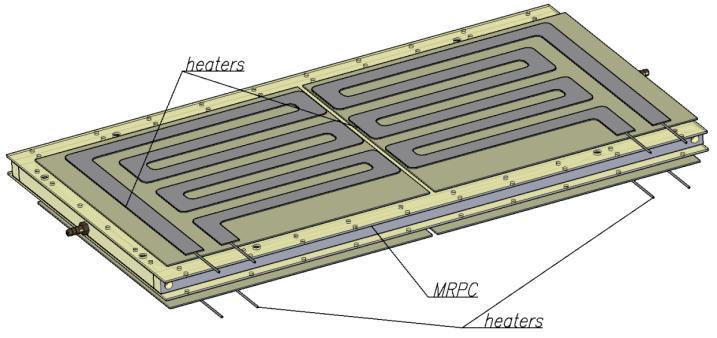}
\caption
{
The heaters on the both sides of MRPC.  
}
\label{fig:fig2}
\end{figure}

%A special surface heaters were designed to  warm a chamber.
A thermal stabilization system of the chamber consists of two pairs of heaters and a microcontroller. 
Each pair of heaters covers half of the chamber on the top and on the bottom (see Fig.\ref{fig:fig2}). 
The chamber is wrapped with a 10~mm thick heat-insulating material.
Two digital temperature sensors monitor each pair of heaters.
The microcontroller uses the sensors reading to control the temperature.
To stabilize the temperature a proportional-integral-differential regulator algorithm is implemented in
the microcontroller firmware. The algorithm kept temperature of the chamber with the accuracy $\pm 0.3^\circ$C.

\section{Front-end electronics}
\label{fee}

\begin{figure}[hbtp]
\centering
\includegraphics*[width=70mm,height=50mm]{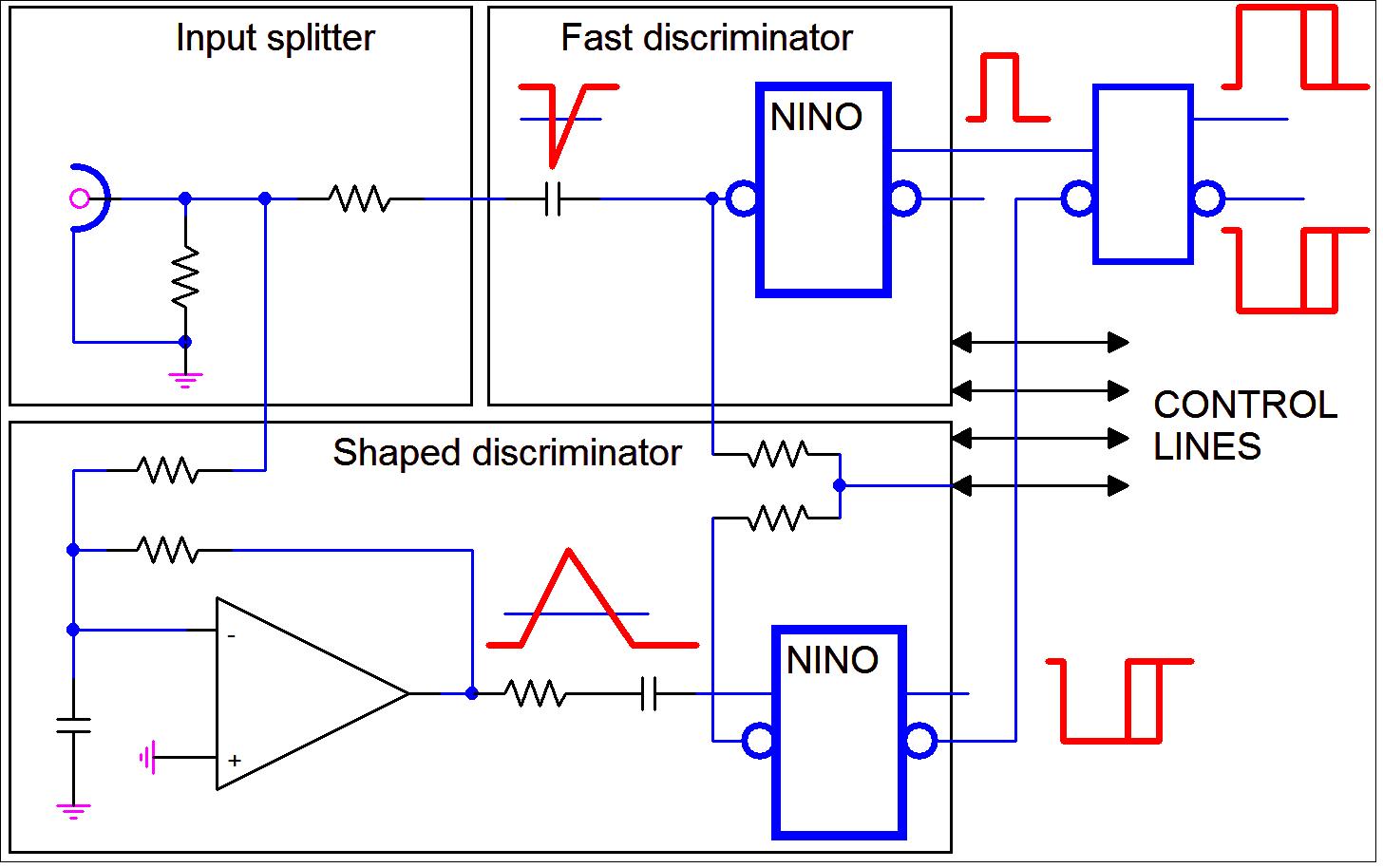}
\caption
{A diagram of the MRPC front-end electronics consisting of input splitter, fast and shaped 
discriminators, control unit and LVDS driver.
}
\label{fig:fig3}
\end{figure}

The signals  from the both sides of each strip of MRPC are fed to an  
amplifier-discriminator using 50$\Omega$ coaxial cables 30~cm long. 
A 32-channels front-end electronics board is based on the NINO chip \cite{nino}.
The output signal of the NINO amplifier-discriminator is the time-over-threshold (TOT) pulse. 
The leading edge of the pulse corresponds to the time of the hit. 
Pulse width is proportional to the input signal charge \cite{nino, daq_hades}.  
A diagram of the MRPC front-end electronics (FEE) is shown in Fig.\ref{fig:fig3}.
FEE consists of input splitter, fast and shaped discriminators, control unit and LVDS driver.
 
A two-threshold discriminator structure is chosen to ensure exact correspondence 
between the input charge and the pulse duration. The leading and trailing edges of the pulse at the output 
of the board are formed by the fast discriminator (FD) and by the shaped discriminator (SD), respectively. 
The scheme provides an accurate time reference of leading edge with time jitter less than 10 ps. 
The correspondence of the pulse duration to the input charge is found to be as accurate as 0.3\% 
(for triangular pulses from a generator).

The output LVDS signals from the amplifier-discriminator are
fed  to 32-channel VME time-to-digital converter TDC32VL with time-sampling of 25 ps \cite{TDC_afi}.
Power supply, threshold settings, stretch time settings and hysteresis settings
are controlled by the U-40 VME module \cite{U40_afi}.

\section{U70 test beam}
\label{U70}

\begin{figure}[hbtp]
\centering
\includegraphics*[width=70mm,height=30mm]{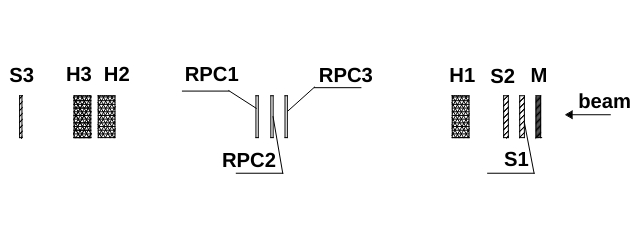}
\caption
{Schematic view of the  MRPC test setup (not in scale). $H1-H3$ are the hodoscopes, $S1-S3$ are the trigger 
scintillation counters, $M$ is the  monitor counter, $RPC1-3$  are the MRPCs under test.
}
\label{fig:fig4}
\end{figure}

The MRPC test setup is schematically shown in Fig.\ref{fig:fig4}. 
The data were taken using a muon beam at U-70 accelerator in Protvino  \cite{U70_status}. The muons were originated from the interaction of the circulating proton beam with the internal target placed in the accelerator vacuum chamber.
No momentum selection was applied to the beam. The setup consisted 
of three stations of hodoscopes ($H1-H3$), three trigger scintillation counters ($S1-S3$),  monitor scintillation counter ($M$) and three MRPCs. The trigger counters 
and the hodoscopes covered 19$\times$19 cm$^2$ of the MRPC sensitive area.  The hodoscopes were used for muon track reconstruction 
with a spatial resolution of 0.6 mm.
 Muons with large scattering angles were filtered out. The monitor counter measured muon 
flux on the MRPC during the data taking.
Duration of the beam spill was about 1 s with the repetition rate of 0.1 Hz. 
Before data taking the ~~MRPCs were operated for 10 days at 15 kV and exposed to radiation at the counting rate $\sim$800 Hz/cm$^2$ for 6 days.

The distance between the MRPCs was 5 cm. Variations in the muon time of flight was neglected in the time resolution calculation. 
Eight strips in each chamber were used for the analysis.  
 
Time differences between the MRPC$_{1-3}$ and  the trigger
(reference scintillation counter) signal were measured for each strip. 
Since the strip was read at both ends, the mean time  is independent of the hit position along the strip direction. 
Three distributions of the time difference between three independent pairs of the
chambers were obtained to eliminate the effect of the trigger signal time resolution.
The widths of these distributions were used to
calculate the individual time resolution of each MRPC by solving the system of linear equations. There is a MRPC time shift (slewing) 
depending on the signal amplitude.  FEE based on the use of NINO chip \cite{nino} measures the input charge and encodes it into the
width of the LVDS pulse; this information is used to correct for
slewing. We used time-width correction function (TOT method) to eliminate MRPC  slewing. However, our procedure was different from that described in 
ref.\cite{nim_cern}. 

To find out time-width
correction function for a particular MRPC we were compelled to use the time response of neighbouring  MRPC as a reference time. Obviously the reference time requires a time-width correction also. Hence we developed an iterative procedure for finding time-width correction for the each strip of each MRPC. This procedure includes a sub-process of a time-shift correction for the strip when corresponding time-width correction function was modified. 

The procedure of the time-width correction is the following.
For each event we did selection of hit strips in all chambers using a reconstructed muon track. Each strip was put in compliance with  a (neighboring)  reference MRPC. The time difference of the hit strip and corresponding reference MRPC was used to calculate the time-width correction function of that strip.
In this way, 24 time-width correction functions were calculated for three tested  MRPCs. The response time of each strip was corrected for the corresponding   dependence.  After that we did a equalization for the mean time responses of
the strips. For that we solved a system of linear equations with the fixed mean time response of one (arbitrary selected) strip.
The time shifts for other strips were found as a solution of this system.
The final response time  for each strip was calculated by the subtraction of the calculated  time shift.
The iteration procedure was stopped when variation in  correction functions and time shifts become less than $\sim$1 ps. Usually, such procedure required no more than 5 steps.

The data acquisition system recorded a time stamp for each event. This information was used
to study a behavior of MRPCs efficiency and the time resolution versus exposition time.

\section{Results}
\label{results}

\begin{figure}[hbtp]
\centering
\includegraphics*[width=60mm,height=55mm]{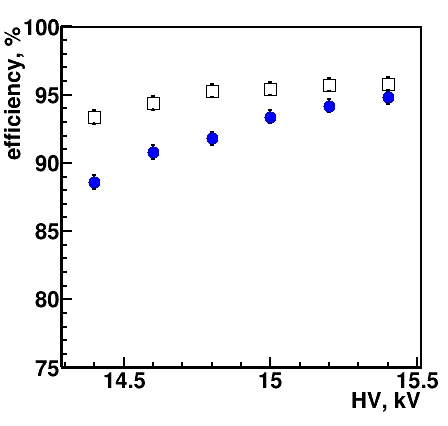}
\caption
{The efficiency of the MRPC as a function of the applied voltage at the counting rate 
$\sim$450 Hz/cm$^2$ and the temperature 25$^\circ$C. The open squares and full circles represent the results obtained with small trigger counters and using large acceptance setup, respectively.}
\label{fig:fig5}
\end{figure}

\begin{figure}[hbtp]
\centering
\includegraphics*[width=60mm,height=55mm]{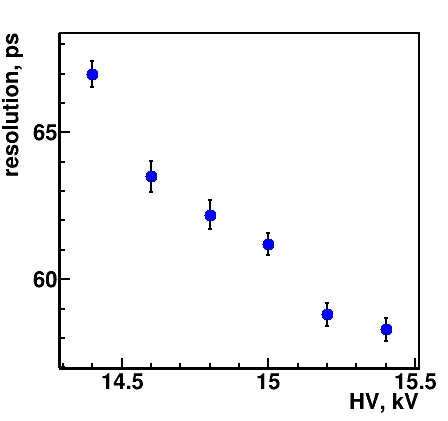}
\caption
{The time resolution of the MRPC as a function of the applied voltage at the counting 
rate $\sim$450 Hz/cm$^2$ and the temperature 25$^\circ$C.}
\label{fig:fig6}
\end{figure}

The efficiency and the time resolution  of the MRPC as a function of applied voltage 
are shown by the full circles in Fig.\ref{fig:fig5} and Fig.\ref{fig:fig6}, respectively. 
The gas mixture was 98\% of freon C$_2$F$_4$H$_2$  and 2\% of SF$_6$.
The results were  
obtained at the temperature 25$^\circ$C and the counting rate $\sim$450 Hz/cm$^2$.
The MRPC efficiency is calculated as a ratio of strip hit number to the reconstructed muon
tracks crossing.  A voltage plateau is not observed, because 
sometimes the track was incorrectly associated with the strip due to  multiple scattering.
A crosstalk amplitude in such strip increases with the voltage increase what results in an efficiency grow. The measured dark count rate does not exceed $\sim$1 Hz/cm$^2$ for all three MRPCs. Therefore, it cannot influence on the efficiency growing. The efficiency is $\sim$95\% and the time resolution is 58 ps at the voltage of 15.4 kV.
The working point of high voltage 
for MRPCs was  obtained using  scintillation counters with
the sizes of 1$\times$1 cm$^2$ by the HV scan performed before main measurements.
The results obtained with small trigger counters are shown in  Fig.\ref{fig:fig5} by open
squares. The effect of the multiple scattering biases the efficiency towards lower values with respect to the measurement done with small scintillators, where no position matching between the triggered track and the hit on the MRPC is required. The  crosstalk partially recovers this effect at higher HV and this causes a missing plateau, that was instead present when using the small scintillators.

\begin{figure}[hbtp]
\centering
\includegraphics*[width=60mm,height=55mm]{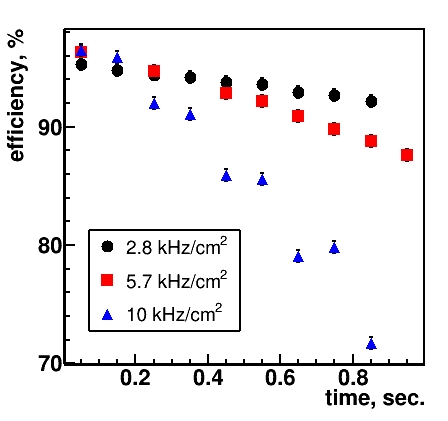}
\caption
{Dependence of the MRPC efficiency on an exposition time at a temperature 30$^\circ$C 
for mean counting rates of 2.8, 5.7 and 10 kHz/cm$^2$.
}
\label{fig:fig7}
\end{figure} 

\begin{figure}[hbtp]
\centering
\includegraphics*[width=60mm,height=55mm]{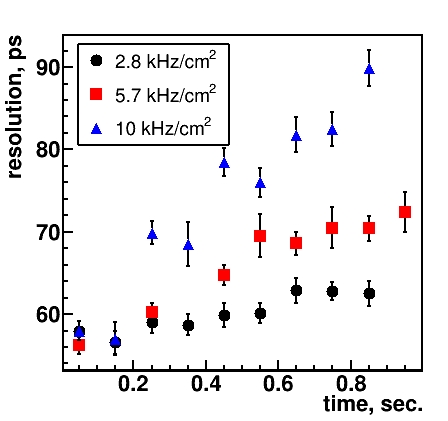}
\caption
{Dependence of the MRPC time resolution on the exposition time at a temperature 30$^\circ$C
for mean counting rates of 2.8, 5.7 and 10 kHz/cm$^2$.
}
\label{fig:fig8}
\end{figure} 

\begin{figure}[hbtp]
\centering
\includegraphics*[width=60mm,height=55mm]{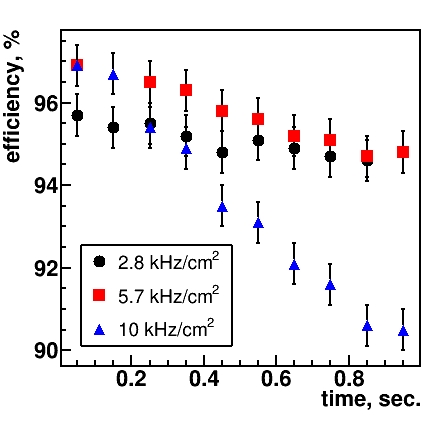}
\caption
{Dependence of the MRPC efficiency  on the exposition time at a temperature 40$^\circ$C
at mean counting rates of 2.8, 5.7 and 10 kHz/cm$^2$.
}
\label{fig:fig9}
\end{figure} 

\begin{figure}[hbtp]
\centering
\includegraphics*[width=60mm,height=55mm]{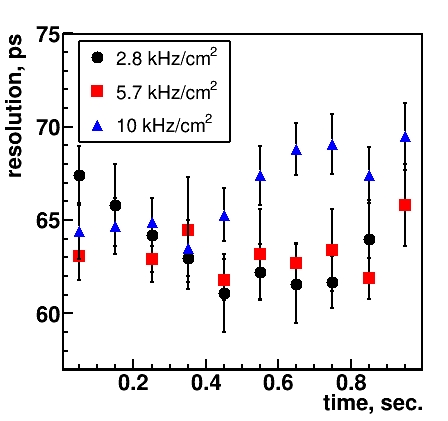}
\caption
{Dependence of the MRPC  time resolution  on the exposition time at a temperature 40$^\circ$C
at mean counting rates of 2.8, 5.7 and 10 kHz/cm$^2$.
}
\label{fig:fig10}
\end{figure}

The dependence of the MRPC efficiency and time resolution  on the exposition time at the counting 
rates from 2.8 to 10 kHz/cm$^2$ are shown 
in Fig.\ref{fig:fig7}- Fig.\ref{fig:fig10}, respectively. 
It is worth to note that the beam intensity profile was flat during the accelerator spill. 
The results were averaged for the MRPC voltage 15.0 and 15.2 kV. 
Such procedure can be used, because the MRPC efficiency differs on less than 1\%
for these two high voltages (see Fig.\ref{fig:fig5}).     
Fig.\ref{fig:fig7} and  Fig.\ref{fig:fig8} demonstrate the changes 
in the MRPC characteristics obtained at the temperature 30$^\circ$C at 
different counting rates.  
At the mean counting rate of  2.8 kHz/cm$^2$ the efficiency decreases $\sim$5\%, while the time resolution degrades from 58 ps to 64 ps to the end of the beam spill.
These changes become more significant for higher rates. 
The efficiency decreases by $\sim$10\% and more than 30\% in one second for 
the counting rates  of 5.7 and 10 kHz/cm$^2$, respectively. 
The time resolution degrades by $\sim$20 ps and $\sim$40 ps  in one second at 5.7 and 10 kHz/cm$^2$, respectively.

\begin{figure}[hbtp]
\centering
\includegraphics*[width=60mm,height=55mm]{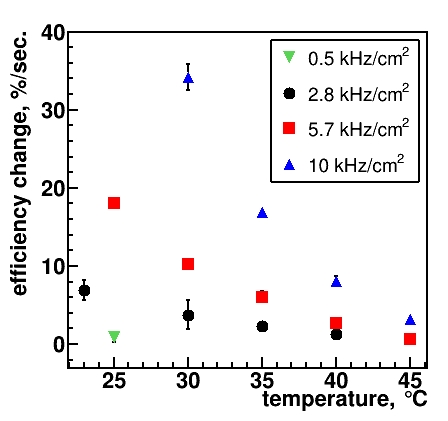}
\caption
{A pace of the efficiency at the beginning of the spill versus the 
MRPC temperature at different counting rates. 
}
\label{fig:fig11}
\end{figure} 

\begin{figure}[hbtp]
\centering
\includegraphics*[width=60mm,height=55mm]{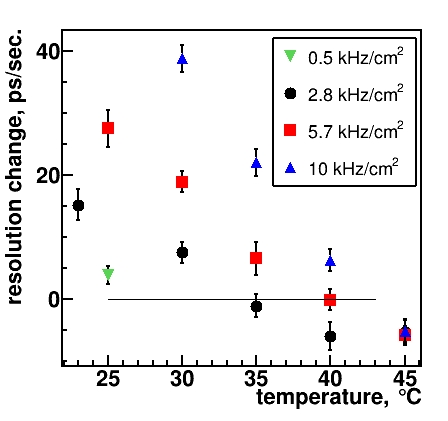}
\caption
{An initial pace of the time resolution change dependence on the MRPC
temperature at different counting rates. 
}
\label{fig:fig12}
\end{figure}

Dependencies of the efficiency and the time resolution of the MRPC 
 at the temperature 40$^\circ$C on the exposition time for different counting
rates are shown in Fig.\ref{fig:fig9} and  Fig.\ref{fig:fig10}, respectively.
At this temperature efficiency drop is less than 2\% for the counting rates up 
to  $\sim$6 kHz/cm$^2$ which is consistent with the results obtained for the  MRPCs made of low resistivity glass \cite{china1}.
Comparison of the data presented in Fig.\ref{fig:fig7} and Fig.\ref{fig:fig9}
demonstrates significant efficiency increase when the temperature rises to 40$^\circ$C.

The time resolution remains almost constant with the exposition time for the counting
rate of 5.7 kHz/cm$^2$, while it decreases at higher
rate (10 kHz/cm$^2$). However, the maximal change in the time resolution does not 
exceed 5 ps, which is significant improvement relative to the one at 30$^\circ$C (see Fig.\ref{fig:fig8}).

 It is obviously important to have a stable TOF system.
Therefore, the optimal operating temperature at certain counting rate 
should correspond to minimal variation of the MRPC characteristics.
To estimate the pace of efficiency variation an approximation function $A+B\cdot e^{-\alpha\cdot t}$, was used. Here $t$ is the exposition time, while $\alpha$, $A$ and $B$ are the 
parameters.
%taken in the form $A+B\cdot e^{-\alpha\cdot t}$, while 
A linear function was  used to fit the data on the time resolution. 
Dependence of the pace of efficiency and the time resolution changes on the
temperature are  shown in Fig.\ref{fig:fig11} and Fig.\ref{fig:fig12}, respectively.
The measurements at temperature 25$^\circ$C and counting rate $\sim$450 Hz/cm$^2$ were also added.
As one can see from Fig.\ref{fig:fig11} the efficiency of the MRPC at the temperature  40$^\circ$C  and 45$^\circ$C 
does not change during the spill at the counting rates less than 3 kHz/cm$^2$ and  6 kHz/cm$^2$, respectively.
The stability domain of the time resolution fits well to the efficiency one.
%A zero speed of the time resolution change is shifted   
%on the temperature scale by about five degrees down (see Fig.\ref{fig:fig14}). 
%In practice, such a shift is not significant, since  the speed curve of the efficiency 
%change becomes flat when approaching zero, and typically the spill duration does not exceed 2 s. 
Negative pace of the time resolution change means that the time resolution at the beginning of the beam spill improves. 
%Such behavior is due to the MRPC overheating  due to low counting rate between the beam spills. 
This effect is seen in  Fig.\ref{fig:fig10} at the counting rate of 2.8 kHz/cm$^2$.  
%The operating temperature range required for a particular counting rate can be selected
%using the obtained experimental temperature dependencies presented in Fig.\ref{fig:fig13}
%and Fig.\ref{fig:fig14}.

\begin{figure}[hbtp]
\centering
\includegraphics*[width=60mm,height=55mm]{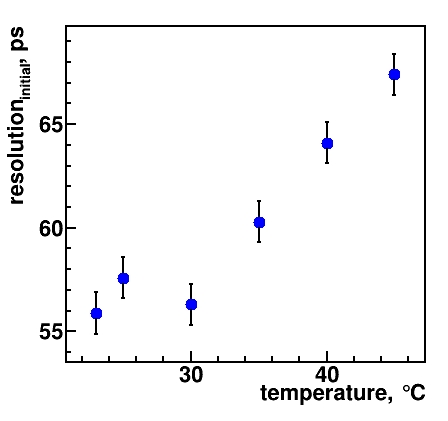}
\caption
{The MRPC time resolution at the beginning of beam spill versus temperature.}
\label{fig:fig13}
\end{figure} 

The detector time resolution depends on the radiation load and on the temperature. At low counting rates there is a dependence on the temperature only.
In addition to the pace of  the time resolution change, 
the detector temperature affects the  time resolution value at the beginning of the 
beam spill (initial resolution), before which the MRPC was almost no radiation load. 
The temperature dependence of an initial time resolution is shown in Fig.\ref{fig:fig13}.
The initial time resolution degrades by $\sim$10 ps with temperature increase from 22$^\circ$C to 45$^\circ$C.
However, such   worsening is not critical, since the time resolution of the MRPC at 
a temperature of 25$^\circ$C is about 57 ps.

In the temperature range from 20$^\circ$C to 50$^\circ$C  
the resistivity of the glass can be parameterized as
\begin{eqnarray}
\label{Rglass}
\rho = e^{8.1-0.114\cdot T}\times 10^{10} \Omega\cdot cm,
\end{eqnarray}
where T is a temperature given in $^\circ$C \cite{gapienko1}.
The  volume  resistance of the electrodes of the MRPC operating   
at temperatures from 40$^\circ$C to 45$^\circ$C at the counting rates up to 
6 kHz/cm$^2$, will be 4.4-5.9 times less than at a temperature of 25$^\circ$C,
at which the maximal MRPCs load does not exceed 0.5 kHz/cm$^2$.
Therefore,   the dependence of the maximal possible counting rate on the chamber and 
the MRPC operating temperature is  non linear.

\section{Conclusions}
\label{concl}

A twelve gap resistive plate chamber equipped with surface heaters to increase
the counting rate capability was designed.
The MRPC counting rate capability was studied in the temperature range 
25$^\circ$C -  45$^\circ$C  using the muon beam at U-70 accelerator  \cite{U70_status}. 
It was shown that the MRPC at the temperature 40$^\circ$C can sustain the charged particle flux 
up to 6 kHz/cm$^2$ with a time resolution about 65 ps and the efficiency higher than 95\%. 
It was found that  these characteristics do not deteriorate within few seconds after beginning of the beam spill.  

The inner part of the TOF system of the BM@N spectrometer 
will be subjected to the charged particle flux up to 5-6 kHz/cm$^2$. 
Such load goes far beyond the capability of the conventional float-glass MRPCs technology.
The designed  MRPC heated to 40-45$^\circ$C satisfies 
the conditions of the BM@N experiment for the $Au+Au$ central collisions \cite{bmn1,bmn2}.

The presented results open a broad prospect for this TOF technique to deal with high particle flux.
 
\vspace{0.5cm}

{\bf Acknowledgments}\\

The authors thank the U-70 crew for providing a good condition
of the experiment. They are  grateful  to V.A.~Gapienko  for useful 
discussions.



%% References


%\bibliographystyle{elsarticle-num}



\end{document}